\documentclass[prb,twocolumn,showpacs,preprintnumbers,amsmath,aps]{revtex4}
\usepackage{graphicx,hyperref}
\usepackage{xcolor}
\usepackage{bm}
\usepackage{subfigure}

\def\nbC{{\mathchoice {\setbox0=\hbox{$\displaystyle\rm C$}%
\hbox{\hbox to0pt{\kern0.4\wd0\vrule height0.9\ht0\hss}\box0}}
{\setbox0=\hbox{$\textstyle\rm C$}\hbox{\hbox
to0pt{\kern0.4\wd0\vrule height0.9\ht0\hss}\box0}}
{\setbox0=\hbox{$\scriptstyle\rm C$}\hbox{\hbox
to0pt{\kern0.4\wd0\vrule height0.9\ht0\hss}\box0}}
{\setbox0=\hbox{$\scriptscriptstyle\rm C$}\hbox{\hbox
to0pt{\kern0.4\wd0\vrule height0.9\ht0\hss}\box0}}}}
\def\nbQ{{\mathchoice {\setbox0=\hbox{$\displaystyle\rm
Q$}\hbox{\raise
0.15\ht0\hbox to0pt{\kern0.4\wd0\vrule height0.8\ht0\hss}\box0}}
{\setbox0=\hbox{$\textstyle\rm Q$}\hbox{\raise
0.15\ht0\hbox to0pt{\kern0.4\wd0\vrule height0.8\ht0\hss}\box0}}
{\setbox0=\hbox{$\scriptstyle\rm Q$}\hbox{\raise
0.15\ht0\hbox to0pt{\kern0.4\wd0\vrule height0.7\ht0\hss}\box0}}
{\setbox0=\hbox{$\scriptscriptstyle\rm Q$}\hbox{\raise
0.15\ht0\hbox to0pt{\kern0.4\wd0\vrule height0.7\ht0\hss}\box0}}}}
\def\nbT{{\mathchoice {\setbox0=\hbox{$\displaystyle\rm
T$}\hbox{\hbox to0pt{\kern0.3\wd0\vrule height0.9\ht0\hss}\box0}}
{\setbox0=\hbox{$\textstyle\rm T$}\hbox{\hbox
to0pt{\kern0.3\wd0\vrule height0.9\ht0\hss}\box0}}
{\setbox0=\hbox{$\scriptstyle\rm T$}\hbox{\hbox
to0pt{\kern0.3\wd0\vrule height0.9\ht0\hss}\box0}}
{\setbox0=\hbox{$\scriptscriptstyle\rm T$}\hbox{\hbox
to0pt{\kern0.3\wd0\vrule height0.9\ht0\hss}\box0}}}}
\def\nbS{{\mathchoice
{\setbox0=\hbox{$\displaystyle     \rm S$}\hbox{\raise0.5\ht0%
\hbox to0pt{\kern0.35\wd0\vrule height0.45\ht0\hss}\hbox
to0pt{\kern0.55\wd0\vrule height0.5\ht0\hss}\box0}}
{\setbox0=\hbox{$\textstyle        \rm S$}\hbox{\raise0.5\ht0%
\hbox to0pt{\kern0.35\wd0\vrule height0.45\ht0\hss}\hbox
to0pt{\kern0.55\wd0\vrule height0.5\ht0\hss}\box0}}
{\setbox0=\hbox{$\scriptstyle      \rm S$}\hbox{\raise0.5\ht0%
\hboxto0pt{\kern0.35\wd0\vrule height0.45\ht0\hss}\raise0.05\ht0%
\hbox to0pt{\kern0.5\wd0\vrule height0.45\ht0\hss}\box0}}
{\setbox0=\hbox{$\scriptscriptstyle\rm S$}\hbox{\raise0.5\ht0%
\hboxto0pt{\kern0.4\wd0\vrule height0.45\ht0\hss}\raise0.05\ht0%
\hbox to0pt{\kern0.55\wd0\vrule height0.45\ht0\hss}\box0}}}}
\def\nbZ{{\mathchoice {\hbox{$\sf\textstyle Z\kern-0.4em Z$}}
{\hbox{$\sf\textstyle Z\kern-0.4em Z$}}
{\hbox{$\sf\scriptstyle Z\kern-0.3em Z$}}
{\hbox{$\sf\scriptscriptstyle Z\kern-0.2em Z$}}}}

\begin{document}

\title{Comment on ``Evidence for Supersymmetry in the Random-Field Ising Model at $D=5$''}

\author{Ivan Balog} \email{balog@ifs.hr}
\affiliation{Institute of Physics, P.O.Box 304, Bijeni\v{c}ka cesta 46, HR-10001 Zagreb, Croatia}

\author{Gilles Tarjus} \email{tarjus@lptmc.jussieu.fr}
\affiliation{LPTMC, CNRS-UMR 7600, Sorbonne Universit\'e,
bo\^ite 121, 4 Pl. Jussieu, 75252 Paris cedex 05, France}

\author{Matthieu Tissier} \email{tissier@lptmc.jussieu.fr}
\affiliation{LPTMC, CNRS-UMR 7600, Sorbonne Universit\'e,
bo\^ite 121, 4 Pl. Jussieu, 75252 Paris cedex 05, France}

\date{\today}


\maketitle

In a recent letter,\cite{fytas19} Fytas {\it et al.} study the critical point of the equilibrium random-field Ising model (RFIM) in $D=5$ by means of state-of-art $T=0$ lattice simulations. They show that the underlying supersymmetry (SUSY) of the model,\cite{parisi79} which is clearly broken in $D=4$ and $D=3$, is satisfied to a numerical accuracy of approximately $1\%$ in $D=5$. This result, which complements an earlier simulation study by the same authors\cite{fytas17} on an approximate restoration of the dimensional-reduction (DR) property of the critical scaling in $D=5$, restricts the number of scenarios describing the critical behavior of the RFIM as $D$ is decreased below the upper critical dimension of $6$. Of the two scenarios compatible with their findings, Fytas {\it et al.} suggest that the less likely is the one that we have put forward based on a nonperturbative functional renormalization group (NP-FRG) theory:\cite{tarjus04,tissier06,tissier11,FPbalog}  SUSY and DR apply above a critical dimension $D_{DR}\approx 5.1$ and break down below due to the relevance of collective phenomena known as avalanches (or shocks) that induce singularities (cusps) in the functional dependence of the cumulants of the renormalized random field at the fixed point.\cite{tarjus13} We disagree with the conclusions of Fytas {\it et al.} on two key points: 1) our scenario does not imply that the correction-to-scaling exponent $\omega$ should be small and scale as $\omega\sim D_{DR}-D\approx 0.1$ in $D=5$ but is instead perfectly compatible with the large value $\omega=0.66(+15/-13)$ found numerically\cite{fytas17,fytas19} and 2)  a $1\%$ accuracy is not sufficient to claim that SUSY is restored in $5 D$. We substantiate our two comments below.

1) The functional character of the NP-FRG previously allowed us to unravel the peculiar way by which the new fixed point where both SUSY and DR are broken emerges from the collapse of two SUSY-DR fixed points (one stable and one unstable) in $D=D_{DR}$.\cite{FPbalog} This emergence takes place through a boundary-layer mechanism that leads to a discontinuity in the lowest irrelevant eigenvalue exactly in $d_{DR}$ and to a rapid increase of $\omega$ below $D_{DR}$, in the form of a square-root singularity. It allows one to escape the curse of a small correction-to-scaling exponent in $D=5$: We indeed find from the solution of the NP-FRG equations (derived and discussed in Refs. [\onlinecite{tissier11,FPbalog}]) in $D=5$ a  value $\omega\approx 0.65$ that is fully compatible with the numerical result of Fytas {\it et al.}.\cite{fytas17,fytas19} (The other exponents are found to be $\eta \approx 0.044$, $\bar\eta\approx 0.048$, $\nu\approx 0.627$, in good agreement with the simulation results, $0.052$, $0.058$, $0.626$,\cite{fytas17,fytas19} respectively.) We display in Fig. 1 the $D$ dependence of the smallest irrelevant eigenvalue around the fixed point in the vicinity of $D_{DR}$. This eigenvalue is equal to $\omega$ below $D_{DR}$ (whereas, due to the ``cuspy'' functional nature of the associated eigenfunction, it is not observable in the correction to scaling of usual observables above $D_{DR}$). 

2) We show in the inset of Fig. 1 the very small but {\it nonzero} violation of the SUSY Ward identity displayed by NP-FRG fixed-point functions in $D=5$, which is always smaller than $1\%$ (and can be even smaller in other observables) as a result of the proximity of $D_{DR}$.

\begin{figure}[tbp]
\includegraphics[width=.8 \linewidth]{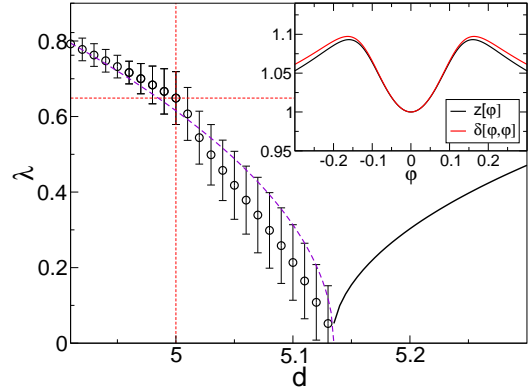}
\caption{Lowest irrelevant eigenvalue of the stability matrix of the RFIM fixed point numerically obtained from our NP-FRG approach in the vicinity of the critical dimension $D_{DR} \approx 5.13$.  This eigenvalue is discontinuous in $D_{DR}$ and has square-root singularities on approaching $D_{DR}$ from both sides. (As the numerical resolution of the functional RG equations is done by discretizing the field arguments on a grid,\cite{tissier11,FPbalog} it is very difficult to properly approach $D_{DR}$ from below and the curve there is just a guide for the eye.) Inset: Test of a SUSY Ward identity at the NP-FRG fixed point in $D=5$: SUSY implies that the second cumulant of the renormalized random field $\delta_*(\varphi,\varphi)$ is equal to the field renormalization function $z_*(\varphi)$.\cite{tissier11} The observed violation is by less than $1\%$, as found for a related Ward identity in [\onlinecite{fytas19}].} 
\label{Fig_eigenvalues}
\end{figure}

\begin{acknowledgments}
IB acknowledges the support of the Croatian Science Foundation Project No. IP-2016-6-3347 and the QuantiXLie Centre of Excellence, a project cofinanced by the Croatian Government and European Union through the European Regional Development Fund - the Competitiveness and Cohesion Operational Programme (Grant KK.01.1.1.01.0004). IB also thanks the LPTMC for its hospitality and the CNRS for funding during the spring of 2019.
\end{acknowledgments}

\end{document}